\documentclass[12pt]{article}
\usepackage[english]{babel}

\usepackage[T1]{fontenc}
\usepackage[utf8]{inputenc}
\usepackage[english]{babel}
\usepackage{cite}

\usepackage{amsthm}
\usepackage{csquotes}

\usepackage[left=2cm,right=2cm,
top=2cm,bottom=2cm,bindingoffset=0cm]{geometry}

\usepackage{hyperref}

\usepackage{indentfirst} 

\usepackage{fancybox} 

\usepackage{latexsym} 
\usepackage{amsfonts} 
\usepackage{amssymb}  
\usepackage{amsmath}  
\usepackage{amstext}  
\usepackage{bm}       

\usepackage[table]{xcolor} 

\usepackage{booktabs} 
\usepackage{tabularx} 

\newtheorem{fact}{Proposition}
\newtheorem*{fact1}{Proposition 1}
\newtheorem*{fact2}{Proposition 2}
\newtheorem*{fact3}{Proposition 3}

\newtheorem{alg}{Algorithm}

\def\bkR{{\rm I\kern-.17em R}}

\usepackage{float}
\usepackage{epsfig}
\usepackage{graphicx}
\graphicspath{{./img/}}

\usepackage{multirow}

\usepackage{hyperref}

\usepackage{indentfirst} 

\usepackage{fancybox} 

\usepackage{latexsym} 
\usepackage{amsfonts} 
\usepackage{amssymb}  
\usepackage{amsmath}  
\usepackage{amstext}  
\usepackage{bm}       
\usepackage{array}

\usepackage{graphicx} 

\usepackage[table]{xcolor} 

\usepackage{booktabs} 
\usepackage{tabularx} 
\usepackage{imakeidx}
\makeindex[columns=3, title=Alphabetical Index, intoc]
\usepackage{fancyhdr}
\usepackage{cancel}
\usepackage{tikz}
\usepackage{multicol}
\usepackage{url}
\usepackage{caption}
\usepackage{subcaption}

\usepackage{hyperref}

\DeclareMathOperator{\Cost}{Cost}
\DeclareMathOperator{\Est}{Est}
\DeclareMathOperator{\Var}{Var}
\DeclareMathOperator{\Cov}{Cov}
\DeclareMathOperator{\Grad}{Grad}
\DeclareMathOperator{\mc}{mc}

\usepackage{mathtools}
\mathtoolsset{showonlyrefs=true}

\title{Automatic Adjoint Differentiation for special functions involving expectations}
\author{José Brito\thanks{Department of Mathematics, University of Aveiro, Portugal, josencbrito@ua.pt}, Andrei Goloubentsev\thanks{Barclays, NY, USA and MatLogica, London, UK., andrei.goloubentsev@matlogica.com}, Evgeny Goncharov\thanks{Department of Pure Mathematics and Mathematical Statistics, University of Cambridge, UK., and MatLogica, London, UK., eg555@cam.ac.uk,   evgeny.goncharov@matlogica.com}}
\date{}

\begin{document}
	
	\maketitle
	
	\begin{abstract}
	
	    We explain how to compute gradients of functions of the form $G = \frac{1}{2} \sum_{i=1}^{m} (E y_i - C_i)^2$, which often appear in the calibration of stochastic models, using Automatic Adjoint Differentiation and parallelization. We expand on the work of \cite{aadc} and give faster and easier to implement approaches. We also provide an implementation of our methods and apply the technique to calibrate European options. 
		
		
	\end{abstract}

\tableofcontents

	\section{Introduction}\label{sec1}
	
	
	
	Automatic Adjoint Differentiation (AAD) is a rapidly-growing field with a variety of applications ranging from biology \cite{biology} to machine learning \cite{ml}. It has recently become popular in financial applications for calculating sensitivities (first-order derivatives, the gradient) of functions automatically alongside the computation of the function, see e.g. \cite{savine1, savine2, savine3, ssrn, aadc, adjoints}.
    
    Suppose one has an algorithm for a function $$f: \mathbb R^n_x \to \mathbb R^m_y,$$ or coordinate-wise:
$$
\begin{aligned}
y_1=&f_1(x_1,x_2,\ldots,x_n), \\ 
y_2=&f_2(x_1,x_2,\ldots,x_n), \\ 
&\vdots \\
y_m=&f_m(x_1,x_2,\ldots,x_n). 
\end{aligned}
$$
AAD maps each vector $(\lambda_1, \dots, \lambda_m) \in \mathbb R^m$ to the following set of linear combinations of $\frac{\partial y_i}{\partial x_k}$:
\begin{equation}\label{0711A}
\left \{ \sum_{i=1}^m \lambda_i \frac{\partial y_i}{\partial x_k}, \quad k=1, \ldots , n\right \}.   
\end{equation}
See \cite{ssrn} for an overview of the method and \cite{capriotti} for a comprehensive introduction. 

If the objective function $f$ requires calculating some expectations $Ey_i$ at intermediate steps, then it is generally understood \cite{fries} that one should avoid differentiating the expectations. In \cite[Appendix A]{fries} the author explains how to perform AAD without differentiating expectations in general. 

In this paper we give three explicit algorithms to compute the gradient of a function of the form \begin{equation}\label{eqG} G = \frac{1}{2} \sum_{i=1}^{m} (E y_i - C_i)^2,\end{equation} where $y_i$ are given by a forward algorithm \begin{equation} \label{eq3} F: \mathbb{R}_{(a_1, \dots, a_M | x_1, \dots, x_N)}^{M+N} \rightarrow \mathbb{R}^{m}_{(y_1, \dots, y_m)}\end{equation} with $M$ parameters and $N$ independent random variables, and $C_i$ are some fixed constants. We also assume that we have a reverse algorithm $R$ performing the AAD for F. We require that both $y_i$ and all the partial derivatives $\frac{\partial y_i}{\partial x_j}$ have finite mean and variance\footnote{This condition is satisfied for all natural functions and we use it in Appendix \ref{proofs} to justify the variance bounds.}

Such functions are often used in the calibration of stochastic models. In Section \ref{sec4} we demonstrate how to apply our algorithms to calibrate a set of European options. Functions of this form have been considered before in \cite{aadc} and in fact, our first algorithm is exactly the method given by the authors. We rigorously explain why that approach works and introduce novel approaches that are easier to implement and have a lower computational cost.

We recall the Monte-Carlo method for approximating expectations. Suppose that we want to compute $Ey$ for a random variable $y: \mathbb{R}^N \to \mathbb{R}$. Then we can estimate $Ey$ as
\begin{equation}\label{eqee}
Ey \sim \frac{1}{N_{\mc}} \sum_{j=1}^{N_{\mc}} y(w_j),
\end{equation}
where $w_j, \ j = 1, \dots, N_{\mc}$ in \eqref{eqee} are independent identically distributed random vectors\footnote{We shall use the notation $w_j$ to denote both the random variables and their simulated values (numbers). The distinction should be clear from the context.} with finite mean and variance (simulated sample paths of random values from the set of Monte-Carlo simulations of a given stochastic process) and $N_{\mc}$ is the number of Monte-Carlo paths. 
By the central limit theorem this method displays $\frac{1}{\sqrt{N_{\mc}}}$ convergence. The cost of this operation is \begin{equation} \label{mce} \Cost(Ey) =  N_{\mc}  \Cost(F). \end{equation}

The rest of the paper is organized as follows. In Section \ref{sec2} we explain how to compute the gradient of $G$ using three different methods, compare the computational costs, and discuss how to parallelize the computations. In Section \ref{sec3} we give a real-world example and discuss performance. Finally, in Section \ref{sec4} we use the developed methods to calibrate European options. The necessary mean and variance computations of Section \ref{sec2} are given in Appendix \ref{proofs} and the full implementation code for Section \ref{sec3} is available as a Github project \cite{github}.

\textbf{Acknowlegements.} We are grateful to Evgeny Lakshtanov for the useful advice, discussions, and the suggestion to to write up these ideas. We are also thankful to Matlogica LTD for providing access to the AADC library. 

The work of the first author was supported by Portuguese funds through CIDMA-Center for Research and Development in Mathematics and Applications, and FCT–Fundação para a Ciência e a Tecnologia, within projects UIDB/04106/2020 and UIDP/04106/2020.

\section{Computing the gradient of G}\label{sec2}
	
	To compute the gradient of $G$ we need to calculate all the partial derivatives. Fix a variable $x_k, \ k \in 1, \dots, N$. Then we have: \begin{multline*}
	 \frac{\partial G}{\partial x_k} = \frac{\partial }{\partial x_k} \left(\frac{1}{2} \sum_{i=1}^{m} \left(E y_i - C_i\right)^2  \right) = \\ =\sum_{i=1}^{m} \left(E y_i - C_i\right) \frac{\partial(Ey_i)}{\partial x_k} = \sum_{i=1}^{m} \left(E y_i - C_i\right) E\left(\frac{\partial y_i}{\partial x_k}\right) = \\ = E \left(\sum_{i=1}^{m} \left(E y_i - C_i\right) \frac{\partial y_i}{\partial x_k}\right).   
	\end{multline*} 
	
	We shall now give three methods to compute the gradient, compare the computational costs, and explain how to parallelize the computations.
	
	\subsection{First approach} \label{first}
	
	The first approach follows \cite{aadc}. We can estimate $\frac{\partial G}{\partial x_k}$ by the following expression:
	
	\begin{align}\label{eq4}  \Est_1\left(\frac{\partial G}{\partial x_k}\right) = \frac{1}{N_{\mc}} \sum_{j=1}^{N_{\mc}} \sum_{i=1}^{m} \left( Ey_i - {C_i} \right) \frac{\partial y_i(w_{j})}{\partial x_k}, \end{align} where $w_j$ are independent identically distributed random vectors with finite mean and variance and $N_{\mc}$ is the number of Monte-Carlo paths. This leads to the following algorithm $\Grad_1(G)$ for computing the gradient:
	
	\begin{alg} \label{alg1}
	
	\begin{enumerate}
		
		\item Calculate all the $Ey_i$ using the forward \textit{$F$}. The cost of this operation is $N_{\mc} \Cost(F)$ as in \eqref{mce}.
		
		\item Fix a random vector (Monte-Carlo path) $w_j, j=1, \dots, N_{\mc}$ and apply the reverse $R$ with $\lambda_i = Ey_i - C_i$. By \eqref{0711A} we obtain the following sequence:
		
		\begin{equation} \label{eq8} \left\{ \sum_{i=1}^{m} \left(Ey_i - C_i\right) \frac{\partial y_i(w_j)}{\partial x_k}, \; k = 1, \dots, N \right\}.\end{equation}
		
		Here we need to first apply $F$ and then $R$ so the cost is\footnote{One could also store the result of Step 1 for all $w_j$. Then the computation cost of this step becomes $\Cost(R)$ but that constrains memory usage.} $\Cost(F)+\Cost(R)$.

		\item Finally, for every $k = 1, \dots, N$ average over all paths $w_j, j=1, \dots, N_{\mc}$ to obtain the gradient.
		
	\end{enumerate}
	
	The total cost of the algorithm is given by
	\begin{equation} \label{eq99} \Cost(\Grad_1(G)) = N_{\mc} \left( 2 \Cost (F) + \Cost (R) \right).\end{equation}
	
	\end{alg}
	
	In Appendix \ref{proof1} we prove the following:
	
	\begin{fact} \label{fact1}
	We have

    $$E\left( \Est_1\left(\frac{\partial G}{\partial x_k}\right) \right) = \frac{\partial G}{\partial x_k}$$ and $$N_{mc}\Var \left(\Est_1\left(\frac{\partial G}{\partial x_k}\right) \right)$$ is bounded above by a number independent of $N_{mc}$. 
	\end{fact}
	
Applying the central limit theorem, we see that as $N_{\mc} \rightarrow \infty$,  $\Est_1\left(\frac{\partial G}{\partial x_k}\right)$ converges to $\frac{\partial G}{\partial x_k}$ and the standard deviation $\sqrt{\Var \left(\Est_1\left(\frac{\partial G}{\partial x_k}\right) \right)}$ converges to $0$ at the rate of $\frac{1}{\sqrt{N_{\mc}}}$. So Algorithm \ref{alg1} computes the gradient of $G$ and displays $\frac{1}{\sqrt{N_{\mc}}}$ convergence. Using the calculation in Appendix \ref{proof1} one can find the standard deviation explicitly for a given forward algorithm \eqref{eq3} for $F$, and produce confidence intervals.

	\subsection{Second approach}
	
	It turns out that we can omit the first step in Algorithm \ref{alg1} which leads to lower cost and easier implementation. For that we need to change the estimator\footnote{Here the reason that we use both $w_j$ and $w_{j-1}$ (and hence start from $j=2$) is that we need to use their independence to prove the analog of the first claim in Proposition \ref{fact1}.}:
	
	\begin{align}\label{eq10}  \Est_2 \left(\frac{\partial G}{\partial x_k} \right) = \frac{1}{N_{\mc} - 1} \sum_{j=2}^{N_{\mc}} \sum_{i=1}^{m} \left(y_i(w_{j-1}) - C_i\right) \frac{\partial y_{i}(w_{j})}{\partial x_k}. \end{align}
	
This leads to the following algorithm $\Grad_2(G)$ for computing the gradient:
    
    \begin{alg}\label{alg2}
    
    \begin{enumerate}
		
		\item Fix a random vector (Monte-Carlo path) $w_j, j=2, \dots, N_{\mc}$ and apply the reverse $R$ with\footnote{Here, unless $j=2$, we use the values of $y_i(w_{j-1})$ computed at the previous step.} $\lambda_i = y_i(w_{j-1}) - C_i$. By \eqref{0711A} we obtain the following sequence:
		
		\begin{equation}  \left\{ \sum_{i=1}^{n} (y_i(w_{j-1}) - C_i) \frac{\partial y_i(w_j)}{\partial x_k}, \; k = 1, \dots, N \right\}.\end{equation}
		Here we need to first apply $F$ and then $R$ so the cost is $\Cost(F)+\Cost(R)$.
		
		\item For every $k = 1, \dots, N$ average over all paths $w_j, j=2, \dots, N_{\mc}$ to obtain the gradient.
		
	\end{enumerate}
	
	The total cost of the algorithm is given by
	\begin{equation}  \Cost(\Grad_2(G)) = N_{\mc} \left(\Cost (F) + \Cost (R) \right).\end{equation}
	
	\end{alg}

In Appendix \ref{proof2} we prove the following:

\begin{fact} \label{fact2}
	We have

    $$E\left( \Est_2\left(\frac{\partial G}{\partial x_k}\right) \right) = \frac{\partial G}{\partial x_k}$$ and $$N_{mc}\Var \left(\Est_2\left(\frac{\partial G}{\partial x_k}\right) \right)$$ is bounded above by a number independent of $N_{mc}$. 

	\end{fact}

As in Section \ref{first}, the central limit theorem implies that as $N_{\mc} \rightarrow \infty$,  $\Est_2\left(\frac{\partial G}{\partial x_k}\right)$ converges to $\frac{\partial G}{\partial x_k}$ and the standard deviation $\sqrt{\Var \left(\Est_2\left(\frac{\partial G}{\partial x_k}\right) \right)}$ converges to $0$ at the rate of $\frac{1}{\sqrt{N_{\mc}}}$. So Algorithm \ref{alg2} also computes the gradient of $G$ and displays $\frac{1}{\sqrt{N_{\mc}}}$ convergence. Using the calculation in Appendix \ref{proof1} one can find the standard deviation explicitly for a given forward algorithm \eqref{eq3} for $F$, and produce confidence intervals.

	

So both Algorithm \ref{alg1} and Algorithm \ref{alg2} compute the gradient of any function $G$ of the form \eqref{eqG} but the computational cost of the latter is smaller and the algorithm is easier to implement. Unfortunately, in practice (as we will see in Section \ref{sec32}) $\Var \left(\Est_2\left(\frac{\partial G}{\partial x_k}\right) \right)$ might be much larger than $\Var \left(\Est_1\left(\frac{\partial G}{\partial x_k}\right) \right)$ (for fixed $N_{\mc}$) so the algorithm takes longer to converge. That suggests modifying $\Est_2\left(\frac{\partial G}{\partial x_k}\right)$ to reduce the variance.




	
	
	

\subsection{Third approach} \label{bestapp}

One may use different techniques of variance reduction to modify Algorithm \ref{alg2}. We have found the following to work well. Replacing $y_i(w_{j-1})$ with $$S_i(w_{j-1}) = \frac{1}{j-1} \sum_{m=1}^{j-1}y_i(w_m)$$ in $\Est_2 \left(\frac{\partial G}{\partial x_k} \right)$ we get  \begin{align} \Est_3 \left(\frac{\partial G}{\partial x_k} \right) = \frac{1}{N_{\mc} - 1} \sum_{j=2}^{N_{\mc}} \sum_{i=1}^{m} \left(S_i(w_{j-1}) - C_i\right) \frac{\partial y_{i}(w_{j})}{\partial x_k}. \end{align}
This leads to a third algorithm $\Grad_3(G)$ for computing the gradient:	

\begin{alg}\label{alg3}
    
    \begin{enumerate}
		
		\item Fix a random vector (Monte-Carlo path) $w_j, j=2, \dots, N_{\mc}$ and apply the reverse $R$ with\footnote{Here, unless $j=2$, we use the values of $S_i(w_{j-2})$ and $y_i(w_{j-1})$ from the previous step to compute $S_i(w_{j-1})$.} $\lambda_i = S_i(w_{j-1}) - C_i$. By \eqref{0711A} we obtain the following sequence:
		
		\begin{equation}  \left\{ \sum_{i=1}^{n} (S_i(w_{j-1}) - C_i) \frac{\partial y_i(w_j)}{\partial x_k}, \; k = 1, \dots, N \right\}.\end{equation}
		Here we need to first apply $F$ and then $R$ so the cost is $\Cost(F)+\Cost(R)$.
		
		\item For every $k = 1, \dots, N$ average over all paths $w_j, j=2, \dots, N_{\mc}$ to obtain the gradient.
		
	\end{enumerate}
	
	The total cost of the algorithm is given by
	\begin{equation}  \Cost(\Grad_3(G)) = N_{\mc} \left(\Cost (F) + \Cost (R) \right).\end{equation}
	
	\end{alg}
	
	This algorithm may be regarded as an approximate version of Algorithm \ref{alg1}. Indeed, by (\ref{eqee}) we see that   $S_i(w_{j-1})$ approximates $Ey_i$ at least when $j$ is large. Based on that observation, in Appendix \ref{proof3} we prove the following:
	
	\begin{fact} \label{fact3}
	We have

    $$E\left( \Est_3\left(\frac{\partial G}{\partial x_k}\right) \right) = \frac{\partial G}{\partial x_k}$$ and $$ N_{\mc}^{\frac{2}{3}}\Var \left(\Est_2\left(\frac{\partial G}{\partial x_k}\right) \right)$$ is bounded as $N_{\mc} \to \infty$. 

	\end{fact}
	
	The central limit theorem implies that as $N_{\mc} \rightarrow \infty$,  $\Est_3\left(\frac{\partial G}{\partial x_k}\right)$ converges to $\frac{\partial G}{\partial x_k}$ and the standard deviation $\sqrt{\Var \left(\Est_3\left(\frac{\partial G}{\partial x_k}\right) \right)}$ converges to $0$ at the rate of $\frac{1}{\sqrt[3]{N_{\mc}}}$. So Algorithm \ref{alg3} also computes the gradient of $G$ and displays at least $\frac{1}{\sqrt[3]{N_{\mc}}}$ convergence. Although we can't rigorously prove $\frac{1}{\sqrt{N_{\mc}}}$ convergence nor produce confidence intervals, as is common for variance reduction methods, it turns out that in practice Algorithm \ref{alg2} has $\frac{1}{\sqrt{N_{\mc}}}$ convergence and moreover the variance is similar to that of Algorithm \ref{alg1} (with the same computational cost as Algorithm \ref{alg2}) as we demonstrate in Section \ref{sec32}. So this method combines the best of Algorithms \ref{alg1} and \ref{alg2}.

	\subsection{Parallelizing the computation}\label{sec2.3}
	
	 Suppose that one is given parallelized versions $F_v$ and $R_v$ of $F$ and $R$ respectively that can process $c$ independent sets of input data as in \cite{aadc}. Such $F_v$ and $R_v$ can be obtained from $F$ by using an appropriate AAD tool and $c$ depends on the AAD tool being used. For instance, for an AAD tool tuned to the Intel AVX512 architecture, the natural value of $c$ is $8 \cdot \{ \textit{Number of Cores}\}$. This value is achieved in MatLogica's AADC library \cite{matlogica} that we use to implement the algorithms in Section \ref{sec3}.
	 
	 The costs of parallelized and regular algorithms are related by:
	 	$$
	\begin{aligned}
	\Cost (F_v) = \frac{K_F}{c}  \Cost(F) \\ \Cost (R_v) = \frac{K_R}{c} \Cost(R)
	\end{aligned}
	$$ where $K_F$ and $K_R$ are correction coefficients that reflect the quality of the AAD tool\footnote{Theoretically they should be equal to $1$ but in practice that is never achieved due to intricacies in hardware specifics and software optimization.}.
	
	Since Algorithms \ref{alg1},\ref{alg2}, and \ref{alg3} involve repeated applications of $F$ and $R$, they can be parallelized as well. Replacing $F$ by $F_v$ and $R$ by $R_v$ at each step we obtain parallelized versions $\Grad_1(G)_v$, $\Grad_2(G)_v$, and $\Grad_3(G)_v$ of $\Grad_1(G)$, $\Grad_2(G)$, and $\Grad_3(G)$ respectively. Substituting $\Cost(F_v)$ and $\Cost(R_v)$ into the expressions for $\Cost(\Grad_1(G)), \Cost(\Grad_2(G)), \Cost(\Grad_3(G))$ we obtain the cost estimates:

	$$
	\begin{aligned}
	\Cost(\Grad_1(G)_v) &= \frac{N_{\mc}}{c}  \left( 2 K_F \Cost (F) + K_R \Cost (R) \right)  \\ 	\Cost(\Grad_2(G)_v)= \Cost(\Grad_3(G)_v) &= \frac{N_{\mc}}{c}  \left(  K_F \Cost (F) + K_R \Cost (R) \right)
	\end{aligned}
	$$
	
\section{Implementation}\label{sec3}

In this section we introduce an implementation for Algorithms \ref{alg1}, \ref{alg2}, and \ref{alg3}, and discuss performance.

    \subsection{Coded examples}
    
    In the GitHub project \cite{github} we give coded examples implementing the three methods in C\texttt{++} using MatLogica's AADC Library \cite{matlogica} that is available to try online. The examples take advantage of the parallelization of Section \ref{sec2.3}. To give a real-world example, we think of $G$ as the loss function of a set of European options. Namely, we have the following correspondence:
    \begin{itemize}
        \item $K_i$ - strike prices
        \item $y_i$ - call options $\left( S(T_i) - K_i \right)^+$, where $S(T_i) > 0$ are asset prices at the expiry dates $T_i$ that follow a log-normal distribution\footnote{Since the Monte Carlo paths are random variables following a normal distribution, the exponential functions $S(T_i)$ follow a log-normal distribution.}
        \item $Ey_i$ - expectations of the call options $\left( S(T_i) - K_i \right)^+$ at the expiry dates $T_i$
        \item $C_i$ - observed prices
        \item $x_k$ - knots of interpolation of a piecewise linear curve of volatilities (called $\sigma_k$ in the code)
    \end{itemize}
    
    See \cite{github} for the full listing of the three methods.
	
	\subsection{Performance}\label{sec32}
	
	In the following tables we list the computation times (to compute the gradient) and variances for Algorithms \ref{alg1},\ref{alg2}, and \ref{alg3} for increasing $N_{\mc}$. Below $\Var G_k$ stands for $\Var \left(\Est_i\left(\frac{\partial G}{\partial \sigma_k} \right)\right)$.
	
	\begin{table}[htp!] 
		\centering
		\begin{tabular}{|c||c|c|c|c|c|c|} 
			\hline
			$N_{\mc}$ & \begin{tabular}{@{}c@{}}\textit{Time} \\ $(\mu s)$\end{tabular} & $\Var G_1$ & $\Var G_2$ & $\Var G_3$ & $\Var G_4$ & $\Var G_5$ \\
			\hline
			\hline
			$10^5$ & $34005$ & $1.98307$ & $1.21845$ & $0.645723$ & $0.247607$ & $0.050143$ \\
			\hline
			$10^6$ & $322448$ & $0.199681$ & $0.121639$ & $0.0643522$ & $0.0254993$ & $0.00506854$ \\
			\hline
			$10^7$ & $3368022$ & $0.019912$ & $0.0121564$ & $0.00643526$ & $0.00254602$ & $0.000504794$ \\
			\hline
			$10^8$ & $33873902$ & $0.00199515$ & $0.00121675$ & $0.000643825$ & $0.000254557$ & $0.0000505085$ \\
			\hline
		\end{tabular}
		\caption{Computation times and variances for Algorithm \ref{alg1}.}
		\label{tab1}
	\end{table}


	\begin{table}[htp!] 
		\centering
		\begin{tabular}{|c||c|c|c|c|c|c|}
			\hline
			$N_{\mc}$ & \begin{tabular}{@{}c@{}}\textit{Time} \\ $(\mu s)$\end{tabular} & $\Var G_1$ & $\Var G_2$ & $\Var G_3$ & $\Var G_4$ & $\Var G_5$ \\
			\hline
			\hline
			$10^5$ & $18662$ & $41.1324$ & $25.1392$ & $13.0327$ & $5.19697$ & $1.0067$ \\
			\hline
			$10^6$ & $179363$ & $4.06896$ & $2.50196$ & $1.30933$ & $0.516204$ & $0.0999287$ \\
			\hline
			$10^7$ & $1865660$ & $0.405794$ & $0.249501$ & $0.130456$ & $0.051451$ & $0.00996968$ \\
			\hline
			$10^8$ & $18791130$ & $0.0406675$ & $0.0249804$ & $0.0130731$ & $0.00515374$ & $0.00099882$ \\
			\hline
		\end{tabular}
		\caption{Computation times and variances for Algorithm \ref{alg2}.}
	\end{table}

	\begin{table}[htp!] 
		\centering
		\begin{tabular}{|c||c|c|c|c|c|c|}
			\hline
			$N_{\mc}$ & \begin{tabular}{@{}c@{}}\textit{Time} \\ $(\mu s)$\end{tabular} & $\Var G_1$ & $\Var G_2$ & $\Var G_3$ & $\Var G_4$ & $\Var G_5$ \\
			\hline
			\hline
			$10^5$ & $20004$ & $2.05491$ & $1.25785$ & $0.662883$ & $0.262605$ & $0.0521572$ \\
			\hline
			$10^6$ & $187509$ & $0.200854$ & $0.122493$ & $0.0649201$ & $0.0256051$ & $0.00507159$ \\
			\hline
			$10^7$ & $1925666$ & $0.0198713$ & $0.0121484$ & $0.0064182$ & $0.00253839$ & $0.000503403$ \\
			\hline
			$10^8$ & $19671966$ & $0.00199513$ & $0.001218$ & $0.000644075$ & $0.00025472$ & $0.0000504992$ \\
			\hline
		\end{tabular}
		\caption{Computation times and variances for Algorithm \ref{alg3}.}
	\end{table}
	
	Note that the running times for the second and third approach are significantly faster than those for the first in accordance with our computations. Moreover, we see that $\Var \left(\Est_i\left(\frac{\partial G}{\partial \sigma_k} \right)\right)$ decreases linearly for all three methods, which agrees with Propositions \ref{fact1}, \ref{fact2} (and outperforms the bound of Proposition \ref{fact3} for the third method). Note also that as promised at the end of Section \ref{bestapp} the variances for the third approach are close to those of the first one. We can see clearly that Algorithm \ref{alg3} outperforms the other two. 
	
\section{Calibrating European options}\label{sec4}

    We can now calibrate the set of European options (same as in Section \ref{sec3}) using AADC and a third-party calibration library \cite{lbfgslib} utilizing the LBFGS calibration method \cite{lbfgs}. The objective is to minimize the loss function $G$. We will perform the calibration independently for the three algorithms.  Calculating the gradient of $G$ using the three algorithms yields the following result.
    
    
    
    \begin{table}[htp!]
		\centering
		\begin{tabular}{|c||c|c|c|c|c|c|}
			\hline
			 & \begin{tabular}{@{}c@{}}\textit{Time} \\ $(\mu s)$\end{tabular} & $\frac{\partial G}{\partial \sigma_1}$ & $\frac{\partial G}{\partial \sigma_2}$ & $\frac{\partial G}{\partial \sigma_3}$ & $\frac{\partial G}{\partial \sigma_4}$ & $\frac{\partial G}{\partial \sigma_5}$ \\
			\hline
			\hline
			Algorithm \ref{alg1}, $N_{\mc}=10^6$ & $322448$ & $3031.56$ & $2384.08$ & $1731.89$ & $1095.55$ & $487.355$ \\
			\hline
			Algorithm \ref{alg2}, $N_{\mc}=10^6$ & $179363$ & $3022.03$ & $2377.34$ & $1727.96$ & $1094.61$ & $486.295$ \\
			\hline
			Algorithm \ref{alg3}, $N_{\mc}=10^6$ & $187509$ & $3033.93$ & $2385.75$ & $1733.13$ & $1096.92$ & $487.696$ \\
			\hline
		\end{tabular}
		\caption{Gradient of $G$ using Algorithms \ref{alg1}, \ref{alg2} and \ref{alg3}.}
		\label{tab3}
	\end{table}
	
\begin{figure}[htp!]
        \centering
        \begin{subfigure}{.5\textwidth}
            \includegraphics[width=\linewidth]{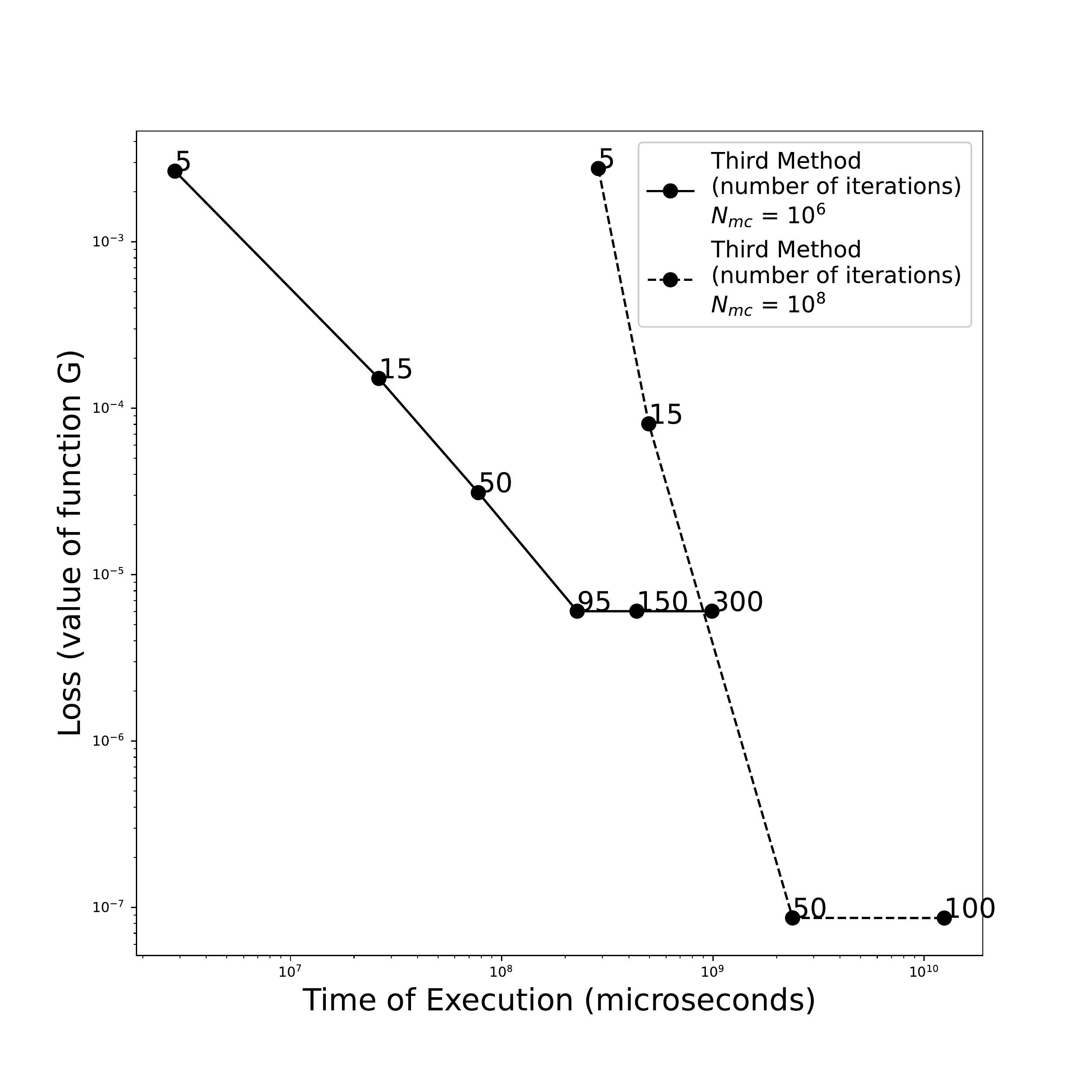}
            \renewcommand{\thesubfigure}{c}
            \caption{Calibrating using Algorithm \ref{alg3} with $N_{\mc}=10^6$ and $N_{\mc}=10^8$.}
            \label{fig:sub5}
        \end{subfigure}
        \renewcommand{\thefigure}{1}
        \caption{Comparison of the three approaches to calibration.}
        \label{fig:test2}
    \end{figure}

	Note that we get similar gradients, and that Algorithms \ref{alg1} and \ref{alg3} converge faster than Algorithm \ref{alg2} in agreement with Section \ref{sec32}. Figure \ref{fig:test2} demonstrates how the three approaches behave in calibrations. We start the graphs (in logarithmic scales) at the 5-th iteration of the calibration algorithm (to see a meaningful difference) and the numbers at the dots correspond to the number of iterations.
	
	

    
    \begin{figure}[htp!]
        \centering
        \begin{subfigure}{.5\textwidth}
            \centering
            \includegraphics[width=\linewidth]{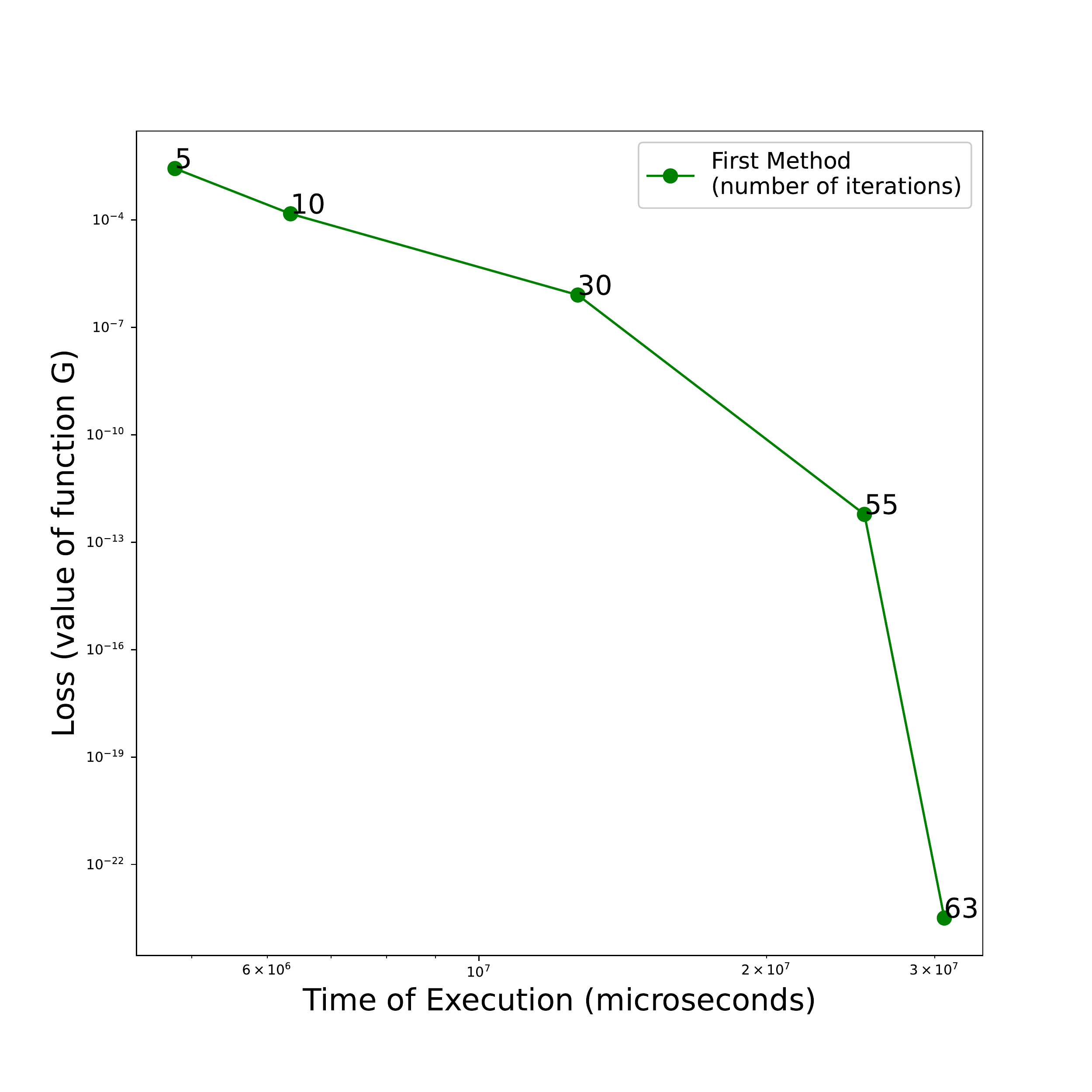}
            \caption{Calibrating using Algorithm \ref{alg1} with $N_{\mc}=10^6$.}
            \label{fig:sub3}
        \end{subfigure}%
        \begin{subfigure}{.5\textwidth}
            \centering
            \includegraphics[width=\linewidth]{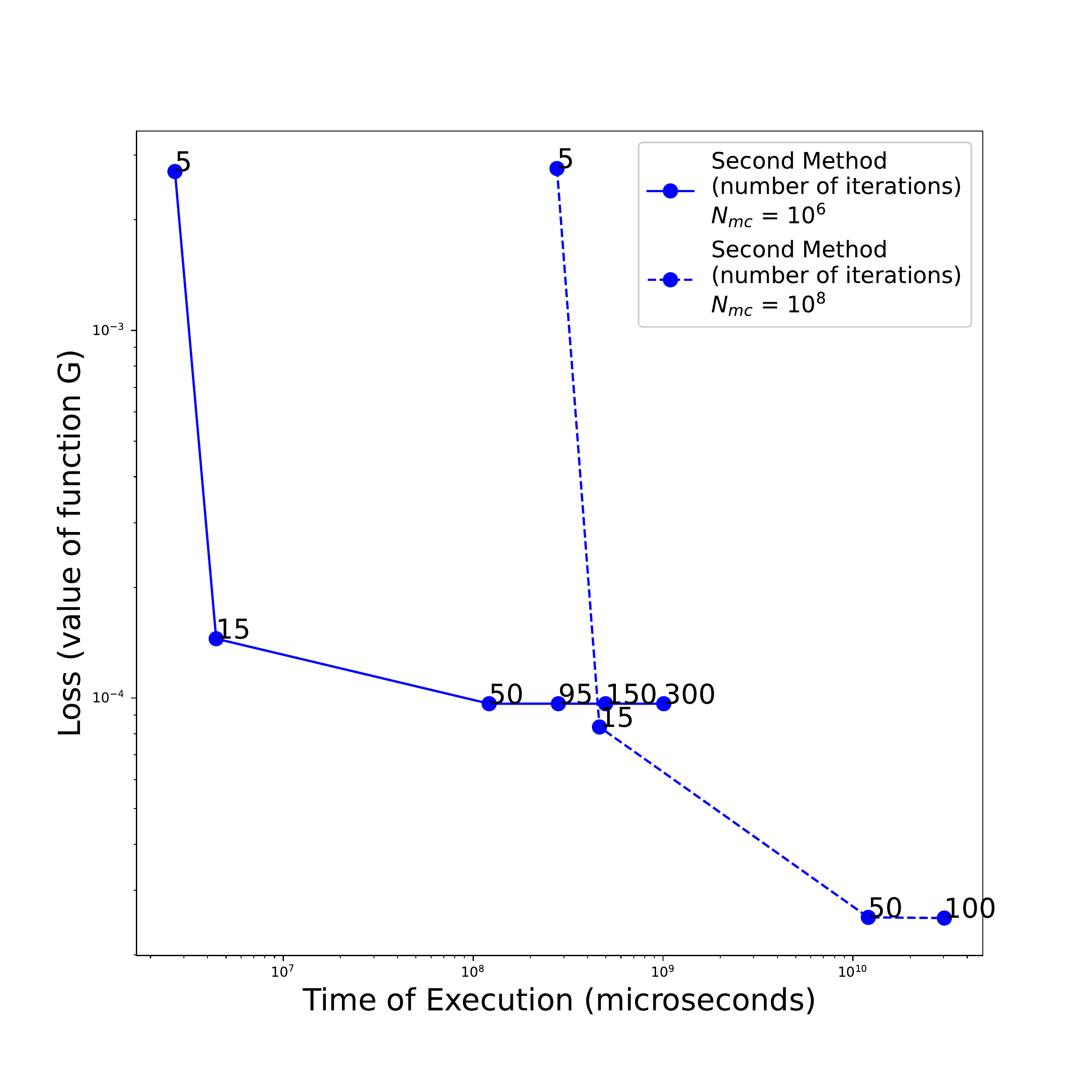}
            \caption{Calibrating using Algorithm \ref{alg2} with $N_{\mc}=10^6$ and $N_{\mc}=10^8$.}
            \label{fig:sub4}
        \end{subfigure}
    \end{figure}
    \clearpage
    
    We can see that in all approaches we successfully minimize the objective function although Algorithms \ref{alg2} and \ref{alg3} require a larger $N_{\mc}$ to calibrate. Calibrating using Algorithm \ref{alg1} is the fastest and Algorithm \ref{alg3} behaves slightly better than Algorithm \ref{alg2} in calibrations.

    

	
	\appendix
	
    \section{Proofs of the Propositions.} \label{proofs}
    
    In this section we prove Propositions \ref{fact1}, \ref{fact2}, and \ref{fact3} justifying that Algorithms \ref{alg1}, \ref{alg2}, and \ref{alg3} compute the gradient of $G$. We show that Algorithms \ref{alg1} and \ref{alg2} display $\frac{1}{\sqrt{N_{\mc}}}$ convergence and Algorithm \ref{alg3} displays $\frac{1}{\sqrt[3]{N_{\mc}}}$ convergence at least asymptotically (although as we have seen in Section \ref{sec32} in practice the convergence is also at the $\frac{1}{\sqrt{N_{\mc}}}$ rate).
    
    \subsection{Proof of Proposition \ref{fact1}} \label{proof1}
    
    \begin{fact1} 
	We have

    $$E\left( \Est_1\left(\frac{\partial G}{\partial x_k}\right) \right) = \frac{\partial G}{\partial x_k}$$ and $$N_{mc}\Var \left(\Est_1\left(\frac{\partial G}{\partial x_k}\right) \right)$$ is bounded above by a number independent of $N_{mc}$. 
	\end{fact1}
	
	\begin{proof}
	The first claim follows from the following computation:
	 \begin{multline*} E\left( \Est_1\left(\frac{\partial G}{\partial x_k}\right) \right) =
	E\left(\frac{1}{N_{\mc}} \sum_{j=1}^{N_{\mc}} \sum_{i=1}^{m} \left( Ey_i - {C_i} \right) \frac{\partial y_i(w_{j})}{\partial x_k} \right) = \\ = \frac{1}{N_{\mc}} \sum_{j=1}^{N_{\mc}} E\left( \sum_{i=1}^{m} \left( Ey_i - {C_i} \right) \frac{\partial y_i(w_{j})}{\partial x_k} \right) = E \left(\sum_{i=1}^{m} \left(E y_i - C_i\right) \frac{\partial y_i}{\partial x_k}\right) = \frac{\partial G}{\partial x_k}
	\end{multline*} where the penultimate  equality is due to the fact that $w_j$ are identically distributed.
	
	For the second claim, rewrite $\Var \left(\Est_1\left(\frac{\partial G}{\partial x_k}\right) \right)$ as follows:
	\begin{multline*} \Var \left(\Est_1\left(\frac{\partial G}{\partial x_k}\right) \right) = \Cov \left(\Est_1\left(\frac{\partial G}{\partial x_k}\right), \Est_1\left(\frac{\partial G}{\partial x_k}\right) \right) = \\ = \Cov \left( \frac{1}{N_{\mc}} \sum_{j=1}^{N_{\mc}} \left( \sum_{i=1}^{m} \left( Ey_i - {C_i} \right) \frac{\partial y_i(w_{j})}{\partial x_k} \right), \frac{1}{N_{\mc}} \sum_{j=1}^{N_{\mc}} \left( \sum_{i=1}^{m} \left( Ey_i - {C_i} \right) \frac{\partial y_i(w_{j})}{\partial x_k} \right)\right) = \\ = \sum_{1 \leq t, l \leq N_{\mc}} \frac{1}{N_{\mc}^2} \Cov\left( \sum_{i=1}^{m} \left( Ey_i - {C_i} \right) \frac{\partial y_i(w_{t})}{\partial x_k}, \sum_{i=1}^{m} \left( Ey_i - {C_i} \right) \frac{\partial y_i(w_{l})}{\partial x_k}  \right) = \\ = \sum_{j=1}^{N_{\mc}} \frac{1}{N_{\mc}^2} \Var\left( \sum_{i=1}^{m} \left( Ey_i - {C_i} \right) \frac{\partial y_i(w_{j})}{\partial x_k} \right) = \frac{1}{N_{\mc}} \Var\left( \sum_{i=1}^{m} \left( Ey_i - {C_i} \right) \frac{\partial y_i (w_1)}{\partial x_k} \right).
	\end{multline*}
	Here we have used that covariance is linear in each argument and that $w_j$ are independent and identically distributed. It suffices to show that $$\Var\left( \sum_{i=1}^{m} \left( Ey_i - {C_i} \right) \frac{\partial y_i}{\partial x_k} \right)$$ is finite.
	
	Indeed, we have 
	\begin{multline*}
	\Var\left( \sum_{i=1}^{m} \left( Ey_i - {C_i} \right) \frac{\partial y_i}{\partial x_k} \right) = \\ =  \sum_{1 \leq t, l \leq m} \Cov \left( \left( Ey_t - {C_t} \right) \frac{\partial y_t}{\partial x_k}, \left( Ey_l - {C_l} \right) \frac{\partial y_l}{\partial x_k} \right) = \\ = \sum_{1 \leq t, l \leq m} \left( Ey_t - {C_t} \right)\left( Ey_l - {C_l} \right)\Cov \left(  \frac{\partial y_t}{\partial x_k},  \frac{\partial y_l}{\partial x_k} \right).
	\end{multline*}
	
	But now $$\left|\Cov \left( \frac{\partial y_t}{\partial x_k}, \frac{\partial y_l}{\partial x_k} \right) \right| \leq \sqrt{\Var\left( \frac{\partial y_t}{\partial x_k} \right)\Var\left( \frac{\partial y_l}{\partial x_k} \right)}$$ which is finite by our assumptions.
	
	\end{proof}

    \subsection{Proof of Proposition \ref{fact2}} \label{proof2}
    
    \begin{fact2} 
		We have

    $$E\left( \Est_2\left(\frac{\partial G}{\partial x_k}\right) \right) = \frac{\partial G}{\partial x_k}$$ and $$N_{mc}\Var \left(\Est_2\left(\frac{\partial G}{\partial x_k}\right) \right)$$ is bounded above by a number independent of $N_{mc}$. 

	\end{fact2}
	
	\begin{proof}
	The first claim follows from the following computation:
	\begin{multline*} E\left( \Est_2\left(\frac{\partial G}{\partial x_k}\right) \right) =
	E\left(\frac{1}{N_{\mc}-1} \sum_{j=2}^{N_{\mc}} \sum_{i=1}^{m} \left(y_i(w_{j-1}) - C_i\right) \frac{\partial y_{i}(w_{j})}{\partial x_k} \right)= \\ = \frac{1}{N_{\mc}-1} \sum_{j=2}^{N_{\mc}} \sum_{i=1}^{m} E\left(y_i(w_{j-1}) - C_i\right) E\left(\frac{\partial y_{i}(w_{j})}{\partial x_k} \right)  = \frac{1}{N_{\mc}-1} \sum_{j=2}^{N_{\mc}} \sum_{i=1}^{m} \left(E y_i - C_i\right) E\left(\frac{\partial y_{i}(w_{j})}{\partial x_k} \right)  = \\ = \frac{1}{N_{\mc}-1} \sum_{j=2}^{N_{\mc}} E \left( \sum_{i=1}^{m} \left(E y_i - C_i\right) \frac{\partial y_{i}(w_{j})}{\partial x_k} \right) = E \left(\sum_{i=1}^{m} \left(E y_i - C_i\right) \frac{\partial y_i}{\partial x_k}\right) = \frac{\partial G}{\partial x_k}
	\end{multline*} where the second equality is because $w_j$ and $w_{j-1}$ are independent and the forth is due to the fact that $w_j$ are identically distributed.  
	
	For the second claim, rewrite $\Var \left(\Est_2\left(\frac{\partial G}{\partial x_k}\right) \right)$ as follows:
		\begin{multline*} \Var \left(\Est_2\left(\frac{\partial G}{\partial x_k}\right) \right) = \Cov \left(\Est_2\left(\frac{\partial G}{\partial x_k}\right), \Est_2\left(\frac{\partial G}{\partial x_k}\right) \right) = \\ = \Cov \left( \frac{1}{N_{\mc} - 1} \sum_{j=2}^{N_{\mc}} \sum_{i=1}^{m} \left(y_i(w_{j-1}) - C_i\right) \frac{\partial y_{i}(w_{j})}{\partial x_k}, \frac{1}{N_{\mc} - 1} \sum_{j=2}^{N_{\mc}} \sum_{i=1}^{m} \left(y_i(w_{j-1}) - C_i\right) \frac{\partial y_{i}(w_{j})}{\partial x_k}\right) = \\ = \sum_{2 \leq t, l \leq N_{\mc}} \frac{1}{(N_{\mc}-1)^2} \Cov\left( \sum_{i=1}^{m} \left(y_i(w_{t-1}) - C_i\right) \frac{\partial y_i(w_{t})}{\partial x_k}, \sum_{i=1}^{m} \left(y_i(w_{l-1}) - C_i\right) \frac{\partial y_i(w_{l})}{\partial x_k}  \right).
	\end{multline*}
	Unlike the proof of \ref{fact1} above, the independence of $w_j$ only implies that $$\Cov\left( \sum_{i=1}^{m} \left(y_i(w_{t-1}) - C_i\right) \frac{\partial y_i(w_{t})}{\partial x_k}, \sum_{i=1}^{m} \left(y_i(w_{l-1}) - C_i\right) \frac{\partial y_i(w_{l})}{\partial x_k}  \right) =0 \text{ if } |m-l| > 1$$ and we can rewrite the sum as 
	
	\begin{multline*}
	\Var \left(\Est_2\left(\frac{\partial G}{\partial x_k}\right) \right)= \sum_{j=2}^{N_{\mc}} \frac{1}{(N_{\mc}-1)^2} \Var \left(\sum_{i=1}^{m} \left(y_i(w_{j-1}) - C_i\right) \frac{\partial y_i(w_{j})}{\partial x_k} \right) + \\ + 2 \sum_{j=2}^{N_{\mc}-1} \frac{1}{(N_{\mc}-1)^2} \Cov \left(\sum_{i=1}^{m} \left(y_i(w_{j-1}) - C_i\right) \frac{\partial y_i(w_{j})}{\partial x_k}, \sum_{i=1}^{m} \left(y_i(w_{j}) - C_i\right) \frac{\partial y_i(w_{j+1})}{\partial x_k} \right).
	\end{multline*}

	Now note that $$\Var \left(\sum_{i=1}^{m} \left(y_i(w_{j-1}) - C_i\right) \frac{\partial y_i(w_{j})}{\partial x_k}\right)$$ is independent of $j$ since $w_j$ are identically distributed and that
	\begin{multline*}
	\left|\Cov \left(\sum_{i=1}^{m} \left(y_i(w_{j-1}) - C_i\right) \frac{\partial y_i(w_{j})}{\partial x_k}, \sum_{i=1}^{m} \left(y_i(w_{j}) - C_i\right) \frac{\partial y_i(w_{j+1})}{\partial x_k} \right)\right| \leq \\ \leq \sqrt{\Var \left(\sum_{i=1}^{m} \left(y_i(w_{j-1}) - C_i\right) \frac{\partial y_i(w_{j})}{\partial x_k} \right) \Var \left(\sum_{i=1}^{m} \left(y_i(w_{j}) - C_i\right) \frac{\partial y_i(w_{j+1})}{\partial x_k} \right)} = \\ = \Var \left(\sum_{i=1}^{m} \left(y_i(w_{j-1}) - C_i\right) \frac{\partial y_i(w_{j})}{\partial x_k} \right).   
	\end{multline*} 
	
	Thus we have 
	\begin{multline*}
	\Var \left(\Est_2\left(\frac{\partial G}{\partial x_k}\right) \right) \leq \\ \leq \left( \frac{1}{N_{\mc}-1} + 2 \cdot  \frac{N_{\mc}-2}{(N_{\mc}-1)^2} \right) \Var \left(\sum_{i=1}^{m} \left(y_i(w_{1}) - C_i\right) \frac{\partial y_i(w_{2})}{\partial x_k} \right) \leq \\ \leq \frac{3}{N_{\mc}-1} \Var \left(\sum_{i=1}^{m} \left(y_i(w_{1}) - C_i\right) \frac{\partial y_i(w_{2})}{\partial x_k} \right)
	\end{multline*}
	and it suffices to show that $$\Var \left(\sum_{i=1}^{m} \left(y_i(w_{1}) - C_i\right) \frac{\partial y_i(w_{2})}{\partial x_k} \right)$$ is finite.
	
	Indeed, we have $$\Var \left(\sum_{i=1}^{m} \left(y_i(w_{1}) - C_i\right) \frac{\partial y_i(w_{2})}{\partial x_k} \right) =  \sum_{1 \leq t, l \leq m} \Cov  \left( \left(y_t(w_{1}) - C_t\right) \frac{\partial y_t(w_{2})}{\partial x_k},  \left(y_l(w_{1}) - C_l\right) \frac{\partial y_l(w_{2})}{\partial x_k} \right).$$
	
	But now 
	\begin{multline*}
	\left|\Cov  \left( \left(y_t(w_{1}) - C_t\right) \frac{\partial y_t(w_{2})}{\partial x_k},  \left(y_l(w_{1}) - C_l\right) \frac{\partial y_l(w_{2})}{\partial x_k} \right) \right| \leq \\ \leq \sqrt{\Var\left( \left(y_t(w_{1}) - C_t\right) \frac{\partial y_t(w_{2})}{\partial x_k} \right)\Var\left( \left(y_l(w_{1}) - C_l\right) \frac{\partial y_l(w_{2})}{\partial x_k} \right)}  
	\end{multline*}
	and it is enough to show that $$\Var\left( \left(y_i(w_{1}) - C_i\right) \frac{\partial y_i(w_{2})}{\partial x_k} \right)$$ is finite for all $1 \leq i \leq m$. That follows from our assumptions and the fact that for two independent variables $X$ and $Y$ one has $$\Var(XY) = \Var(X)\Var(Y) + \Var(X)E(Y)^2 + \Var(Y)E(X)^2.$$
	
	\end{proof}
	
	\subsection{Proof of Proposition \ref{fact3}} \label{proof3}
	
	\begin{fact3} 
 We have

    $$E\left( \Est_3\left(\frac{\partial G}{\partial x_k}\right) \right) = \frac{\partial G}{\partial x_k}$$ and $$ N_{\mc}^{\frac{2}{3}}\Var \left(\Est_2\left(\frac{\partial G}{\partial x_k}\right) \right)$$ is bounded as $N_{\mc} \to \infty$. 

	\end{fact3}
	
	\begin{proof}
	The first claim follows as in the proof of Proposition \ref{fact2} noting that $$E\left(S_i(w_{j-1})\right) = E\left( \frac{1}{j-1} \sum_{m=1}^{j-1}y_i(w_m) \right) = \frac{1}{j-1} \sum_{m=1}^{j-1}E y_i = E y_i.$$ 	
	For the second claim, rewrite $\Var \left(\Est_3\left(\frac{\partial G}{\partial x_k}\right) \right)$ as follows\footnote{Below $\lfloor x \rfloor$ stands for the largest integer less than or equal to $x$.}:
		\begin{multline*} \Var \left(\Est_3\left(\frac{\partial G}{\partial x_k}\right) \right) =  \Var \left( \frac{1}{N_{\mc} - 1} \sum_{j=2}^{N_{\mc}} \sum_{i=1}^{m} \left(S_i(w_{j-1}) - C_i\right) \frac{\partial y_{i}(w_{j})}{\partial x_k}\right) = \\ = \Var \left( \frac{1}{N_{\mc} - 1} \sum_{j=2}^{\lfloor \sqrt[3]{N_{\mc}} \rfloor} \sum_{i=1}^{m} \left(S_i(w_{j-1}) - C_i\right) \frac{\partial y_{i}(w_{j})}{\partial x_k} + \frac{1}{N_{\mc} - 1} \sum_{j=\lfloor \sqrt[3]{N_{\mc}} \rfloor + 1}^{N_{\mc}} \sum_{i=1}^{m} \left(S_i(w_{j-1}) - C_i\right) \frac{\partial y_{i}(w_{j})}{\partial x_k}\right) = \\ = \Var \left( \frac{1}{N_{\mc} - 1} \sum_{j=2}^{\lfloor \sqrt[3]{N_{\mc}} \rfloor} \sum_{i=1}^{m} \left(S_i(w_{j-1}) - C_i\right) \frac{\partial y_{i}(w_{j})}{\partial x_k}\right) + \\ + 2 \Cov \left( \frac{1}{N_{\mc} - 1} \sum_{j=2}^{\lfloor \sqrt[3]{N_{\mc}} \rfloor} \sum_{i=1}^{m} \left(S_i(w_{j-1}) - C_i\right) \frac{\partial y_{i}(w_{j})}{\partial x_k} , \frac{1}{N_{\mc} - 1} \sum_{j=\lfloor \sqrt[3]{N_{\mc}} \rfloor + 1}^{N_{\mc}} \sum_{i=1}^{m} \left(S_i(w_{j-1}) - C_i\right) \frac{\partial y_{i}(w_{j})}{\partial x_k}\right) + \\ + \Var \left(  \frac{1}{N_{\mc} - 1} \sum_{j=\lfloor \sqrt[3]{N_{\mc}} \rfloor + 1}^{N_{\mc}} \sum_{i=1}^{m} \left(S_i(w_{j-1}) - C_i\right) \frac{\partial y_{i}(w_{j})}{\partial x_k}\right).
	\end{multline*}

We now show that all three terms are $\mathcal{O}\left(\frac{1}{N_{\mc}^{\frac{2}{3}}} \right)$ as $N_{\mc} \to \infty$. Indeed, expanding the first term by linearity we get:

\begin{multline*}
\Var \left( \frac{1}{N_{\mc} - 1} \sum_{j=2}^{\lfloor \sqrt[3]{N_{\mc}} \rfloor} \sum_{i=1}^{m} \left(S_i(w_{j-1}) - C_i\right) \frac{\partial y_{i}(w_{j})}{\partial x_k}\right) \leq \\ \leq \frac{{\lfloor \sqrt[3]{N_{\mc}} \rfloor}^2}{\left(N_{\mc}-1 \right)^2} \max_{2 \leq t, l \leq \lfloor \sqrt[3]{N_{\mc}} \rfloor} \left \{ \Cov \left( \sum_{i=1}^{m} \left(S_i(w_{t-1}) - C_i\right) \frac{\partial y_{i}(w_{t})}{\partial x_k}, \sum_{i=1}^{m} \left(S_i(w_{l-1}) - C_i\right) \frac{\partial y_{i}(w_{l})}{\partial x_k}\right) \right \} \leq \\ \leq \frac{{\lfloor \sqrt[3]{N_{\mc}} \rfloor}^2}{\left(N_{\mc}-1 \right)^2} \max_{2 \leq t, l \leq \lfloor \sqrt[3]{N_{\mc}} \rfloor} \left \{ \sqrt{\Var \left( \sum_{i=1}^{m} \left(S_i(w_{t-1}) - C_i\right) \frac{\partial y_{i}(w_{t})}{\partial x_k} \right) \Var \left( \sum_{i=1}^{m} \left(S_i(w_{l-1}) - C_i\right) \frac{\partial y_{i}(w_{l})}{\partial x_k} \right)} \right \}.
\end{multline*}

By a similar argument, $\Var \left( S_i (w_{j-1})  \right) $ is finite for any $j$ and using this all the $$\Var \left( \sum_{i=1}^{m} \left(S_i(w_{l-1}) - C_i\right) \frac{\partial y_{i}(w_{l})}{\partial x_k} \right)$$ are finite as well as in the proof of Proposition \ref{fact2}.  So the first term is $\mathcal{O}\left(\frac{1}{N_{\mc}^{\frac{4}{3}}} \right)$ as $N_{\mc} \to \infty$. 

Similarly, the second term is bounded above by $$\frac{2{\lfloor \sqrt[3]{N_{\mc}} \rfloor} N_{\mc}}{\left(N_{\mc}-1 \right)^2} \max_{2 \leq t \leq \lfloor \sqrt[3]{N_{\mc}} \rfloor < l \leq N_{\mc}} \left \{ \sqrt{\Var \left( \sum_{i=1}^{m} \left(S_i(w_{t-1}) - C_i\right) \frac{\partial y_{i}(w_{t})}{\partial x_k} \right) \Var \left( \sum_{i=1}^{m} \left(S_i(w_{l-1}) - C_i\right) \frac{\partial y_{i}(w_{l})}{\partial x_k} \right)} \right \}$$ and so it is $\mathcal{O}\left(\frac{1}{N_{\mc}^{\frac{2}{3}}} \right)$ as $N_{\mc} \to \infty$.

Finally, we use (\ref{eqee}) to estimate the third term. As $N_{\mc} \to \infty$ we have:

\begin{multline*}
    \Var \left(  \frac{1}{N_{\mc} - 1} \sum_{j=\lfloor \sqrt[3]{N_{\mc}} \rfloor + 1}^{N_{\mc}} \sum_{i=1}^{m} \left(S_i(w_{j-1}) - C_i\right) \frac{\partial y_{i}(w_{j})}{\partial x_k}\right) \sim \\ \sim \Var \left(  \frac{1}{N_{\mc} - 1} \sum_{j=\lfloor \sqrt[3]{N_{\mc}} \rfloor + 1}^{N_{\mc}} \sum_{i=1}^{m} \left(Ey_i + \mathcal{O}\left( \frac{1}{\sqrt{j-1}}\right) - C_i\right) \frac{\partial y_{i}(w_{j})}{\partial x_k}\right) \sim \\ \sim \Var \left(  \frac{1}{N_{\mc} - 1} \sum_{j=\lfloor \sqrt[3]{N_{\mc}} \rfloor + 1}^{N_{\mc}} \left( \sum_{i=1}^{m} \left(Ey_i  - C_i\right) \frac{\partial y_{i}(w_{j})}{\partial x_k} + \mathcal{O}\left( \frac{1}{\sqrt{j-1}}\right) \right)\right) \sim \\ \sim \Var \left(  \frac{1}{N_{\mc} - 1} \sum_{j=\lfloor \sqrt[3]{N_{\mc}} \rfloor + 1}^{N_{\mc}}  \sum_{i=1}^{m} \left(Ey_i  - C_i\right) \frac{\partial y_{i}(w_{j})}{\partial x_k}   + \mathcal{O}\left( \frac{1}{\sqrt[3]{N_{\mc}}}\right)\right) \sim \\ \sim \Var \left(\Est_1\left(\frac{\partial G}{\partial x_k}\right) \right) + \mathcal{O}\left( \frac{1}{N_{\mc}^{\frac{2}{3}}}\right)
\end{multline*}
and $\Var \left(\Est_1\left(\frac{\partial G}{\partial x_k}\right) \right)$ is $\mathcal{O}\left( \frac{1}{N_{\mc}}\right)$ as $N_{\mc} \to \infty$ by Proposition \ref{fact1}. So the third term is $\mathcal{O}\left(\frac{1}{N_{\mc}^{\frac{2}{3}}} \right)$ as $N_{\mc} \to \infty$ and we are done.

	\end{proof}

	\bibliography{refs}
	
	\bibliographystyle{plain}
	
\end{document}